\input vanilla.sty
\font\tenbf=cmbx10

\font\ninerm=cmr9

\font\eightrm=cmr8
\font\eightit=cmti8

\TagsOnRight
\hsize=5.0truein
\vsize=7.7truein
\parindent=15pt
\def\ref#1{$^{[#1]}$}
\newcount\fignumber
\fignumber=0
\def\fig#1#2{\advance\fignumber by1
 \midinsert \vskip#1truecm \hsize12truecm
 \baselineskip=15pt \noindent
 {\ninerm {Fig. \the\fignumber} #2}\endinsert}
\def\table#1{\hsize12truecm \baselineskip=15pt \noindent
 {\ninerm {Table 1} #1}}
\def\ref#1{$^{[#1]}$}
\def\sqr#1#2{{\vcenter{\vbox{\hrule height.#2pt
   \hbox{\vrule width.#2pt height#1pt \kern#1pt
   \vrule width.#2pt}\hrule height.#2pt}}}}

\nopagenumbers
\baselineskip=10pt
\leftline{\eightrm International Journal of Modern Physics C\hfil}
\leftline{\eightrm $\copyright$\, World
Scientific Publishing Company \hfil}
\vglue 5pc
\baselineskip=13pt
\centerline{\tenbf PRECISE DETERMINATION OF THE CONDUCTIVITY}
\centerline{\tenbf EXPONENT OF 3D PERCOLATION USING ``PERCOLA''}
\vglue 1cm
\centerline{\eightrm JEAN-MARIE NORMAND$^1$ and HANS J. HERRMANN$^2$}
\baselineskip=12pt
\centerline{$^1$ \eightit CEA, Service de Physique Th\'eorique, CE-Saclay}
\centerline{\eightit F-91191 Gif-sur-Yvette Cedex, FRANCE}
\smallskip
\centerline{$^2$ \eightit P.M.M.H., E.S.P.C.I., 10 rue Vauquelin}
\centerline{\eightit 75231 Paris Cedex 05, France}
\medskip

\vskip 2.0truecm\noindent
\medskip
Using 20 months of CPU time on
our special purpose computer ``Percola'' we determined
the exponent for the normal conductivity at the threshold of
three-dimensional site and bond percolation. The extrapolation analysis
taking into account the first correction to scaling gives
a value of $t/\nu = 2.26\pm 0.04$ and a correction exponent
$\omega$ around 1.4.
\bigskip
\noindent{\bf 1. Introduction}
\medskip 
Percolation~\ref{1} is the simplest model for a random medium
and is applied to describe porous media, gelation,
mixtures of electrical conductors, and many other 
disorder dominated situations.
It is defined on a lattice by occupying 
with probability $p$ randomly and
independently either the sites or the bonds with
conductors. If one considers that
locally current can only flow between conductors that are
nearest neighbors, then  only above a critical value $p_c$ a global
current can flow through the system. 
The global electrical conductivity $\Sigma$ 
rises like $$\Sigma \sim (p-p_c)^{t} \eqno(1) $$ and the accurate
determination of $t$, or more precisely of $t/\nu$ where $\nu$
is the exponent of the correlation length, 
is the aim of the present paper.

In the '80s, after a heated controversy
the critical exponents for the conductivity of 
two-dimensional percolation
were determined quite accurately~\ref{2,3} and various
conjectures~\ref{4} could be ruled out. The three-dimensional
situation remained less clear because of the larger error bars.
Only recently new efforts have been made~\ref{5,6} to obtain more precise
values for $t/\nu$ in three dimensions
using considerably larger computational means.
Gingold and Lobb~\ref{5} find $t/\nu = 2.276\pm 0.012$ and Batrouni et
al.~\ref{6} claim $t/\nu = 2.282\pm 0.005$.

Using the specially constructed processor ``Percola''~\ref{7}, assembled
in Saclay by J.-M. Normand and M. Hajjar between 1984 and 1987,
we did exact calculations on narrow bars of random conductors
(``strip-bar method''), a technique that has given very precise values
in two dimensions~\ref{2} and for the superconducting-conducting
mixture in three dimensions~\ref{8}. For the three-dimensional
conductivity problem treated in this paper
this method a dozen years ago
gave $t/\nu = 2.2\pm 0.1$, however, with much
smaller computational investment~\ref{9}.

In the following section we describe the methods, in the next
section we present our results and analyze them, and finally
we conclude.
\medskip
\noindent{\bf 2. Method}
\medskip
The method we use~\ref{9} is a straightforward generalization
to three dimensions of the technique of Derrida and
Vannimenus~\ref{10}. We calculate the electrical current on a simple
cubic lattice of width $m$ and depth $n$ in the 
horizontal directions, and of length $L $ vertically, growing from
bottom up. Periodic boundary conditions are used in the $y$-direction,
defined in fig.~1. Each bond between nearest neighbors is
randomly conducting, in bond percolation if the bond is occupied
(with probability $p$),
in site percolation if both neighbors are occupied
(each with probability $p$), otherwise it is
insulating. The top and the
bottom planes are two equipotentials (fixed boundary condition). 
One defines by $I_i$ and $V_i$ the current and potential
at the end of each line $i$, $i = 1$ being the index
associated with the top plane. The conductance
matrix $\sigma_{i,j}$, given through $I_i = \sum^{n \cdot m+1}_{i=1}
\sigma_{i,j} V_j$, is updated via $\sigma_{i,j} ' = \sigma_{i,j}
- \sigma_{i,\alpha} \sigma_{\alpha,j} r/(1+\sigma_{\alpha,\alpha} r)$
if a longitudinal bond is added at line $\alpha$ and via
$\sigma_{i,j} ' = \sigma_{i,j} + (\delta_{\alpha,j} - \delta_{\beta,j})
\cdot (\delta_{\alpha,i} - \delta_{\beta,i})/r$ if a transversal
bond is added between lines $\alpha$ and $\beta$ where $r$
is the resistance of the bond. In that way the conductance
is obtained exactly. A very large $L$ (bar configuration)
automatically averages over many configurations leaving
no ambiguity over whether one should average the conductivity, the
resistivity, or functions of them. This is an advantage over methods 
that average over many disjoint cubes~\ref{3,5,6,10}.

\fig{5}{Bar of length $L$, width $m$ and depth $n$. We calculate the
conductivity between the two black planes.}

We will consider, as in~\ref{9}, the cases $m = n-1$ for bond
percolation and $m = n$ for site percolation.
The conductance of the bar per length is then  
$\sigma_{1,1}/L$ which in the limit $L \rightarrow \infty$ 
goes to a value $\sigma_n$. The calculation is performed
for different linear sizes $n$ at the percolation threshold
$p_c = 0.248812$~\ref{11} for bond percolation and 
$p_c = 0.311605$~\ref{11} for site percolation. 
The critical exponent $t/\nu$ is then
extracted by using the finite size scaling:
$$\sigma_n = an^{-t/\nu}(1+bn^{-\omega})\eqno(2)$$
where $a$ and $b$ are non-universal constants and $\omega$ is 
the (first) universal correction to scaling exponent.
Since $\nu$, the correlation length exponent, is known rather
precisely ($\nu = 0.875 \pm 0.003$)~\ref{11} one can determine $t$.

\medskip
\noindent{\bf 3. Results and Analysis}
\medskip
The values obtained with ``Percola'' using a CPU time of
20 months are given in table~1 for the case of bond and of
site percolation. Percola~\ref{7}, a 64-bit floating point processor
runs at 25 MFlops but is specialized to deal with the algorithm
described in the previous section for which it runs faster
than a Cray-XMP mono-processor. The lengths of the bars presented in
table~1 exceed those calculated before~\ref{9} by about
a factor of 1000.

\medskip
\midinsert
\vbox{\tabskip=0pt \offinterlineskip
\def\tablerule{\noalign{\hrule}}
\halign to440pt{\strut#& \vrule#\tabskip=1em plus2em&
 \hfil#\hfil& \vrule#& \hfil#\hfil& \vrule#& \hfil#\hfil& \vrule#&
 \hfil#\hfil& \vrule#& \hfil#\hfil& \vrule#& \hfil#\hfil& \vrule#&
 \hfil#\hfil& \vrule#\tabskip=0pt\cr\tablerule
&&$n$&&$\sigma_n$&&$\Delta \sigma_n [\times 10^{-5}]$
&&$L$&&$\sigma_n$&&$\Delta \sigma_n [\times 10^{-5}]$&&$L$&\cr\tablerule
&&2&&0.0943868&&0.32&&$10^{10}$&&0.1138264&&0.26&&$10^{10}$&\cr\tablerule
&&3&&0.0484353&&0.17&&$10^{10}$&&0.0583394&&0.27&&$10^{10}$&\cr\tablerule
&&4&&0.0288459&&0.10&&$2 \cdot 10^{10}$
&&0.0349430&&0.23&&$1.52 \cdot 10^{10}$&\cr\tablerule
&&5&&0.0187865&&0.17&&$2.5 \cdot 10^{9}$
&&0.0228411&&0.23&&$4.6 \cdot 10^{9}$&\cr\tablerule
&&7&&0.0095380&&0.20&&$10^{9}$&&0.0119717&&0.30&&$10^{9}$&\cr\tablerule
&&8&&&&&&&&0.0091521&&0.12&&$10^{9}$&\cr\tablerule
&&10&&0.0045248&&0.62&&$3.5 \cdot 10^{7}$&&0.0058170
&&0.39&&$10^{8}$&\cr\tablerule
&&12&&0.0030573&&0.27&&$4 \cdot 10^{7}$&&0.0039910
&&0.43&&$2.2 \cdot 10^7$&\cr\tablerule
&&13&&0.0025709&&0.45&&$8.8 \cdot 10^{6}$&&&&&&&\cr\tablerule
&&14&&0.0021858&&0.32&&$2.16 \cdot 10^{7}$&&&&&&&\cr\tablerule
&&15&&0.0018839&&0.57&&$10^{7}$&&0.0024833&&1.32&&$10^{7}$&\cr\tablerule
&&16&&0.0016327&&0.47&&$2 \cdot 10^{6}$&&&&&&&\cr\tablerule}}
\smallskip
\table {Conductance $\sigma_n$ per length for bond (left) and
site (right) percolation for different widths $n$ of the bar. 
We also show the statistical mean square deviation $\Delta \sigma_n$
and the length $L$ of the bar for each case.}
\endinsert

We analyzed the data with various techniques. 
Finding that $t/\nu$ which minimizes
the squared error for fixed $\omega$ (as done in refs.~[2] and
[8]) gives $t/\nu = 2.15 \pm 0.12$ for the sites and
$t/\nu = 2.26 \pm 0.06$ for the bonds. In both cases
$\omega \approx 1.4$ gave the best fits. Since universality
states that these exponents should be the same only the intersection
of the error bars, i.e. $t/\nu = 2.23 \pm 0.06$, should be relevant.

The data for bond percolation are closer to the asymptotic regime as
can be seen from the error bars and the curvature in a plot 
log($\sigma_n n^{t/\nu}$) against $n^{-\omega}$. 
Looking for bond percolation at successive slopes (local
derivative of log($\sigma$) vs log($n$)) against
$n^{-\omega}$ gives the most plausible curve for $\omega$
around 1.4. In that case $t/\nu = 2.25 \pm 0.05$.

Keeping $t/\nu$ fixed and finding the $\omega$ which
minimizes the error (and number of iterations)
in the Runge-Kutta procedure of minimizing the mean square deviation,
gives a dependence $t/\nu$ against $\omega$ as shown
in fig.~2. For $\omega$ around 1.4 we find $t/\nu = 2.26 \pm 0.04$
for bond percolation. Unfortunately, as opposed to the case of the
superconductivity exponent~\ref{8} the prefactor of the site
percolation data has the same sign as for bond percolation.
Therefore the curves for site percolation in fig.~2 do not
intersect with those of bond percolation which does not allow us to
increase the precision by assuming universality and
combining the two data sets.

\fig{10}{Leading exponent $t/\nu$ against the first correction
exponent $\omega$ for bond percolation omitting all sizes
of width less or equal to $n_l$, for $n_l = 2$ (diamonds),
3 (crosses +), 4 (squares) and 5 (crosses x).}

\medskip
\noindent{\bf 4. Conclusion}
\medskip
Although we have obtained extremely precise data for 
the conductance of narrow
systems our error bars for the exponents $t/\nu$ extrapolated
to infinite sizes are not as small as obtained by
other methods~\ref{5,6}. This is due to the fact that we have
seriously taken into account the first corrections to scaling
which seems imperative for the case of small systems.
As with many simulations of high precision,
statistical errors are much smaller than the systematic ones.
In contrast to the two-dimensional case~\ref{2} and
the superconducting case in three dimensions~\ref{8} the relation
between leading and correction exponent is not in opposite
sense for site percolation as compared to bond percolation
giving us a particularly unlucky situation. We conclude 
$t/\nu = 2.26 \pm 0.04$ consistent with previous values
and the Alexander-Orbach conjecture~\ref{4}.

\bigskip
\bigskip
\bigskip
\bigskip
We thank D. Stauffer for help in the evaluation of the data and advice
and for helpful suggestions about the manuscript.
\bigskip
\noindent{\bf References}
\medskip
\item{1.} D. Stauffer and A. Aharony, {\it Introduction to Percolation Theory},
(Taylor and Francis, London, 1994)
\item{2.} J.-M. Normand, H.J. Herrmann and M. Hajjar, J. Stat. Phys.
{\bf 52}, 441 (1988) 
\item{3.}D.J. Frank and C.J. Lobb, J. Phys. Rev. B {\bf
37}, 302 (1988)
\item{4.} S. Alexander and R. Orbach, J. Physique Lett. {\bf 43}, L625
(1982); J. Kert\'esz, J. Phys. A {\bf 16}, L471 (1983)
\item{5.} D.B. Gingold and C.J. Lobb, Phys. Rev. B {\bf 42}, 8220
(1990)
\item{6.} G.G. Batrouni, A. Hansen and B. Larson, preprint
\item{7.} F. Hayot, H.J. Herrmann, J.-M. Normand, P. Farthouat and
M. Mur, J. Comp. Phys. {\bf 64}, 380 (1986);
M. Hajjar, Thesis (Orsay, 1987) and
J.-M. Normand, ACM Conf. Proc., St Malo, {\bf 55} (1988)
\item{8.} J.M. Normand and H.J. Herrmann, Int. J. of Mod. Phys. C
{\bf 1}, 207 (1990)
\item{9.} B. Derrida, D. Stauffer, H.J. Herrmann and J. Vannimenus,
J. Physique Lett. {\bf 44}, L701 (1983)
\item{10.} B. Derrida and J. Vannimenus, J. Phys. A {\bf 15},
L557 (1982)
\item{11.} R.M. Ziff, private communication
\end